# Synchronization as a unifying mechanism for protein folding

Leandro P. Nadaletti*, Beatriz S. L. P. de Lima, and Solange Guimarães
*Civil Engineering Program, Federal University of Rio de Janeiro (UFRJ) - Rio de Janeiro, Brazil*
*E-mail: lnadaletti@coc.ufrj.br**

Different models such as diffusion-collision and nucleation-condensation have been used to unravel how secondary and tertiary structures form during protein folding. However, a simple mechanism based on physical principles that provide an accurate description of kinetics and thermodynamics for such phenomena has not yet been identified. This study introduces the hypothesis that the synchronization of the peptide plane oscillatory movements throughout the backbone must also play a key role in the folding mechanism. Based on that, we draw a parallel between the folding process and the dynamics for a network of coupled oscillators described by the Kuramoto model. The amino acid coupling may explain the mean-field character of the force that propels an amino acid sequence into a structure through self-organization. Thus, the pattern of synchronized cluster formation and growing helps to solve the Levinthal's paradox. Synchronization may also help us to understand the success of homology structural modeling, allosteric effect, and the mechanism responsible for the recognition of odorants by olfactory receptors.

**PACS:** PACS number(s): 87.15.Cc, 05.65.+b, 87.15.A-

Biochemical reactions in cells require that proteins adopt a specific conformation in their native state. Experimental and theoretical studies conducted over recent decades have not fully explained the folding phenomena, despite significant advances in structural biology [1]. However, we know that all the necessary information for proteins to find their native structure is in their amino acid sequence [2]. Conversely, Levinthal's paradox explains that a thorough search for the native conformation is impractical in a reasonable period of time, given the large number of arrangements that the polypeptide chain can adopt [3]. Various models were created to overcome Levinthal's paradox and explain experimental results for different proteins classes. In the Diffusion-Collision (DC) model created by Karplus and Weaver [4], folding proceeds through aggregation stages due to productive diffusion and collision of secondary structure elements (microdomains), independent of the tertiary structure. The discovery that proteins fold without forming intermediate states guided the creation of the Nucleation-Condensation (NC) model [5]. In the NC model, both local interactions, between adjacent residues, and long-range interactions, between distant residues in the primary sequence, are formed in parallel. The folding funnel concept was a shift in the general perspective for describing and analyzing the folding process; under this new perspective, the process was no longer associated with a single, preferred pathway, and proteins fold in parallel along the rough energy landscapes [6]. Although such models enhance our understanding of the problem, a simple mechanism based on physical principles that generally describe folding kinetics and thermodynamics has not yet fully identified. In this study, we hypothesize that the synchronization of the peptide plane oscillatory movements plays a key role in the folding process as a mechanism that may unify the different existing perspectives. We draw a parallel between the folding process and the dynamics for a network of coupled oscillators described by the Kuramoto model to illustrate the inherent concepts of the new proposed model. Using this model, we try to explain how a protein rapidly adopts a specific three-dimensional native structure.

In general, synchronization is the ability for self-organization, wherein two or more self-sustaining dynamic systems adjust their rhythms to adopt a common behavior through a low-intensity mutual influence [7]. Such non-linear phenomena are widespread in nature and are applied in different fields of study such as modeling cardiac cell behavior in biology [8]; the formation of spatiotemporal patterns in chemical reactions for chemistry [9]; and in physics, through Josephson junction arrays [10]. This phenomenon was first recorded in the XVII century by the Dutch scientist Christiaan Huygens, in a report on pendulum clock synchronization [11]. In the 1960s, the topic was revisited by Winfree [12] who examined the behavior of sets of biological oscillators and showed that such systems undergo abrupt transitions to collective synchrony. Kuramoto [13] proposed modifications to Winfree's original formulation that encouraged studies on synchronization of globally coupled phase oscillators. Under this new model, phase evolution for a population of $N$ globally coupled oscillators should obey the following first-order differential equation [14]

$$\frac{d\vartheta_i}{dt} = \omega_i + \frac{K}{N}\sum_{j=1}^{N}\sin(\vartheta_j - \vartheta_i). \quad (1)$$

In the equation, $\vartheta_i$ and $\omega_i$ are the phase and natural frequency, respectively, for the $i^{th}$ oscillator, and $K$ defines the coupling strength value. Eq.(1) shows that synchronization between oscillators fundamentally depends on their natural frequency distributions, level of coupling and signal propagation mode from one oscillator to another. Kuramoto also analyzed the emerging collective behavior from the model as a function of $R$ and $\xi$, which are the coherence for the set of oscillators and global phase, respectively. The mean behavior was described as

$$Re^{i\xi} = \frac{1}{N}\sum_{j=1}^{N} e^{i\vartheta_j}, \quad (2)$$

where $R$ ranges between zero and one. For R=0, the oscillators move incoherently, and for R=1 the oscillators are in complete synchrony. Using these quantities, in a mean-field model, Eq.(1) can be reduced to

$$\frac{d\vartheta_i}{dt} = \omega_i + KR\sin(\xi - \vartheta_i). \quad (3)$$

Kuramoto defined the critical coupling value threshold $Kc$ for synchronization as a function of the probability density associated with the oscillators' natural frequency distribution. When this threshold is exceeded, an analog of a phase transition describes the onset of synchronization.

It is appropriate to represent the proteins through the internal coordinate system and define the main-chain bond lengths, valence angles, and torsion angles for analyzing conformational changes. Due to the bond strength between atoms, a great amount of energy is required to shift bond elongation and bond angle deformation from their equilibrium values [15]. Therefore, we can focus on studying the complex interrelationship between the backbone torsion angles and non-covalent interactions to understand the mechanisms related to structural change. In our model we considered only two degrees of freedom for each amino acid, the known angles of rotation $\phi$ and $\psi$ around the N-C$\alpha$ and C$\alpha$–C bonds, respectively (Fig. 1). It is well known that the peptide bond has a partial double bond character, which restricts free rotation around the N-C bond and causes C$\alpha$CONHC$\alpha$ atoms to arrange in a flat conformation defined as peptide or amide plane.

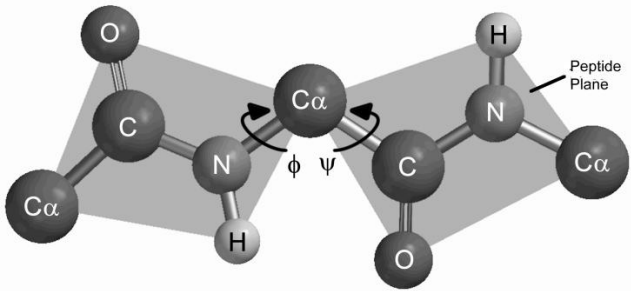

FIG. 1. Torsion angles $\phi$ around the N-C$\alpha$ bond and $\psi$ around the C$\alpha$-C bond. Peptide plane is defined by the C$\alpha$CONHC$\alpha$ atoms positions.

The individual behavior of the peptide plane is expressed by the variation in the angles $\psi$ of the $i^{th}$ amino acid and $\phi$ of the $i+1^{th}$ amino acid, which identifies such a structure as an individual oscillator. This view is supported by the backbone peptide-plane dynamics described in terms of reorientational quasiharmonic modes [16]. Accordingly, we adopted a simplified representation of a protein backbone to describe its conformational dynamics, where the peptide planes were idealized as oscillators held together along the main chain (Fig. 2). Although much more complex, this protein model is comparable to the system of pendulum clocks originally studied by Huygens, for which there are several interpretations regarding the design mode. Pantaleone [17] proposed an interesting variation, which may be tested by assembling a simple device with two pendulum metronomes and a lightweight wooden base supported on two cylinders. A general theoretical analysis of such behavior for a device with several metronomes produced a system of equations to describe phase evolution which is equivalent to the Kuramoto model. A similar model is used to study Josephson junction arrays, a superconducting device operating at high frequency (THz) [18]. A generalization of the original Kuramoto model describes the behavior of series arrays of Josephson junctions [19]. Notably, in the literature, there are several practical cases where the Kuramoto model is directly applied [20].

Baker and Blackburne [21] claimed that molecular structures can behave similar to pendulums in nature. They explained that subcomplexes inside a molecule act as torsion pendulums subject to a harmonic potential. Under certain conditions, the torsional motions may display wide amplitude, resulting in a new oscillation state around a new equilibrium angle. In such situations, the potential energy becomes nonlinear and approximates the potential energy of a real pendulum.

Here, the description of the motion of a peptide plane was approximated by the nonlinear pendulum equation, where it alternates between two limit situations, sometimes behaving as a harmonic oscillator, sometimes as hindered rotor, until it comes to rest in a new equilibrium position. So, a protein may be seen as a chain of peptide planes, acting as oscillators, globally connected with their behavior defined by the Kuramoto model.

The original Kuramoto model assumes a specific form of coupling which is represented by a globally connected graph. Likewise, methods for analysis of protein dynamics, such as parameter-free Gaussian network model (pfGNM) and parameter-free Anisotropy network model (pfANM) assume that all amino acids interact with each other [22]. However, it is possible to adopt a generalized form of Kuramoto model [23] such that the interaction between amino acids can vary over time. In these topologies, instead of a time-independent $K$ from the original proposal, defined as the same for all connections (Eq. 1), a new $K_{ij}(t)$ term, representing the time-varying coupling strengths between amino acids (i) and (j), describes the pattern of interaction observed during the folding process.

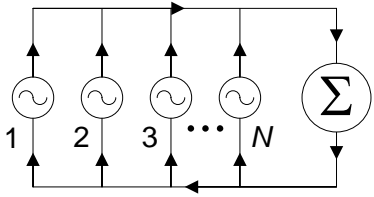

FIG. 2. A schematic model for a protein backbone with $N$ globally coupled peptide planes where each oscillator is driven by the mean field approximation described by Eq. (1).

As can be seen in what follows, correlated motions between distant parts of the proteins were characterized both experimentally and using computer simulations. Experiments performed with the digestive enzyme chymotrypsin showed evidences that vibrational modes at low frequency exist in globular proteins [24]. The Raman spectrum peak, at around $29^{cm-1}$, suggests a dependence upon the conformation adopted, since it disappears with protein denaturation. Throughout the 1970s, Raman scattering experiments identified low frequency modes for a large number of proteins [25]. These results led to the creation of several biophysical models describing the dynamics of macromolecules such as the quasi-continuum model proposed by Chou [26]. Using the concept of low-frequency phonons for proteins, Chou suggested that this internal movement only found in biological macromolecules should result of the collective movement of a large number of atoms. Motions in the folded state are also related to temperatures factors (B Factors) from X-Ray crystallography and to relaxation data from nuclear magnetic resonance (NMR). Specifically, the relationship between relaxation data from NMR and protein dynamics shows that $\psi$ and $\phi$ dihedral angles fluctuations caused by collective protein modes determine NMR order parameters [27]. Moreover, peptide planes dynamics, investigated using Gaussian Axial Fluctuations (GAF) and Residual Dipolar Couplings (RDC), demonstrated a significant anisotropy of the internal motion [28].

Normal mode analysis (NMA) is applied to investigate the vibrational motions that describe the most relevant movements in proteins [29]. As alternative to NMA Molecular Dynamics (MD) simulations are used to better understand the protein dynamics at atomic detail. This technique generates trajectories for each particle in the system by solving Newton's equations of motion [30].

Trajectories from molecular modeling simulations are regarded as non-stationary time series [31]. To analyze the aggregated data in the results of MD simulations, we used a signal processing method, known as the Hilbert-Huang transform (HHT) [32]. This method is divided into two steps. In the first stage, empirical mode decomposition (EMD) is used to decompose the original signal into a set of intrinsic mode functions (IMF). The original signal is expressed by the following equation [32]

$$x(t) = \sum_{k=1}^{N_{imf}} IMF_k(t) + res(t), \quad (4)$$

where $N_{imf}$ is the number of IMFs and res(t) is the expression for the final residue. In the second part, the Hilbert transform is applied to each IMF generated in the previous step. The Hilbert transform of any function g(t) is defined by the equation

$$h(t) = \frac{1}{\pi} P \int_{-\infty}^{+\infty} \frac{g(t)}{t-\tau} d\tau, \quad (5)$$

where $P$ indicates the Cauchy principal value of the singular integral. Based on that definition, h(t) and g(t) form an analytic signal z(t)

$$z(t) = g(t) + ih(t) = a(t)e^{i\delta(t)}, \quad (6)$$

such that $a(t) = (g^2(t) + h^2(t))$ represents the instantaneous amplitude, $\delta(t) = arctan(h(t)/g(t))$ is the instantaneous phase function, and $i = \sqrt{-1}$.

In noisy systems, the phase locking between two periodic oscillators may vary around some average rather than being fixed as described by the equation

$$|m\delta_1 - n\delta_2| < const, \quad (7)$$

Tass [33] proposed an index based on the Shannon entropy to measure statistically the degree of the phase synchrony between two signals. In an attempted to compare the distribution of the instantaneous phase differences to a uniform distribution, the phase coherence value ($\rho$) is defined as

$$\rho = (H_{max} - H)/H_{max}, \quad (8)$$

where $H_{max} = \ln(N_{bin})$ is the maximal entropy. The optimal number of bins was estimated as $N_{bin}=exp(0.626+0.4 \ln(M-1))$, where $M$ is the number of samples. H is the entropy, defined as follows

$$H = -\sum_{k=1}^{N_{bin}} p_k \ln p_k, \quad (9)$$

where $p_k$ is the relative frequency of finding the phase differences within $k$-$th$ bin. Here $\rho = 0$ corresponds to a uniform distribution of the phase differences, i.e., no synchronization is observed, and $\rho = 1$ corresponds to a perfect synchrony.

As a case study, we search for evidence of synchrony within the *cyanovirin*-N protein through the analysis of the backbone torsion angles data collected from the Dynameomics Project database (www.dynameomics.org) [34, 35]. The native state MD simulation that generated the time series were performed at 298K, in a parallel environment developed for analysis and simulation, which is referred to as "in lucem molecular dynamics" (refer to [36] for a detailed description). *Cyanovirin*-N is a mostly beta-sheet protein, consisting of 101 residues (PDB ID: 2EZN) [24]. In practice, only 99 trajectories are available, since the angle $\phi$ for the first residue and the $\psi$ for the last residue are undefined.

For this study, the peptide plane oscillation was approximated using the $\psi$ angle trajectories, since there is a negative correlation between $\psi$ of the $i^{th}$ residue and $\phi$ of the $i+1^{th}$ residue [25]. Those were processed using the EMD algorithm to generate IMFs, as shown in Fig. 3, that were compared to determine the degree of the synchrony between the amino acid pairs.

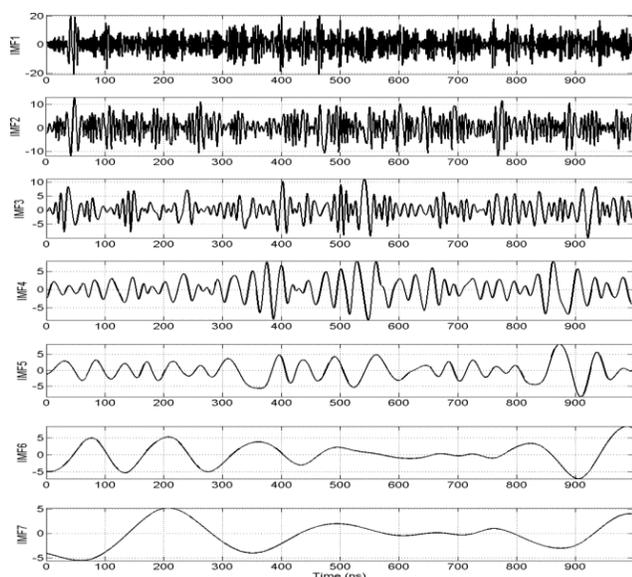

FIG. 3. Signal for 7 IMFs generated by EMD for the $\psi$ trajectory of the residue 54 (PHE) from 0 to 1000 ps.

The analysis of the dynamical interdependence between any two residues of the cyanovirin-N protein was assessed with windows of 1000 ps. The 99 residues were combined two by two in all the possible arrangements (4851 in total) to calculate the phase coherence value ($\rho$) using the Eq. (8). As can be seen in the Fig. 4, over 50% of the pairs present a degree of synchrony greater than 0.6, in the interval from 0 to 1000 ps. Besides that, the largest group has more than 300 pairs with ρ equal to 0.54.

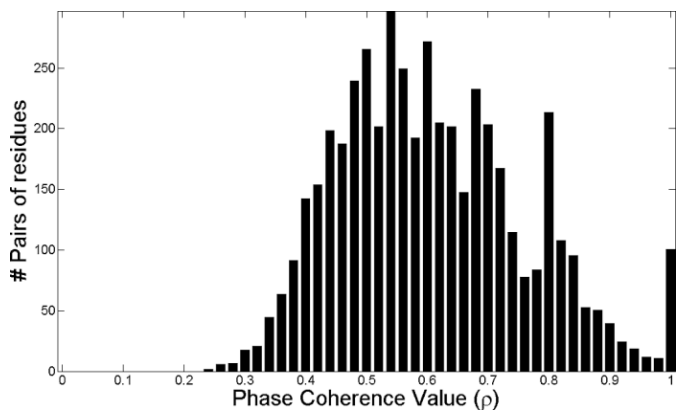

FIG. 4. Histogram detailing the degree of synchrony for 4851 residue pairs, in the interval from 0 to 1000 ps.

The Hilbert-Huang Spectrum (HHS), showing time-energy plots for the low frequency bands of the amino acids PHE(54) and GLU(56), that take part in a small $3_{10}$-helix, can be seen in Fig. 5. In these two graphs, we can observe that during some periods of time the frequencies of different IMFs generate for the amino acid 54 and 56 appear to converge to a common frequency. Furthermore, we notice that this phenomenon occurs simultaneously in both graphs in the time interval from 300 to 400 ps, 700 to 800 ps and 800ps to 900ps.

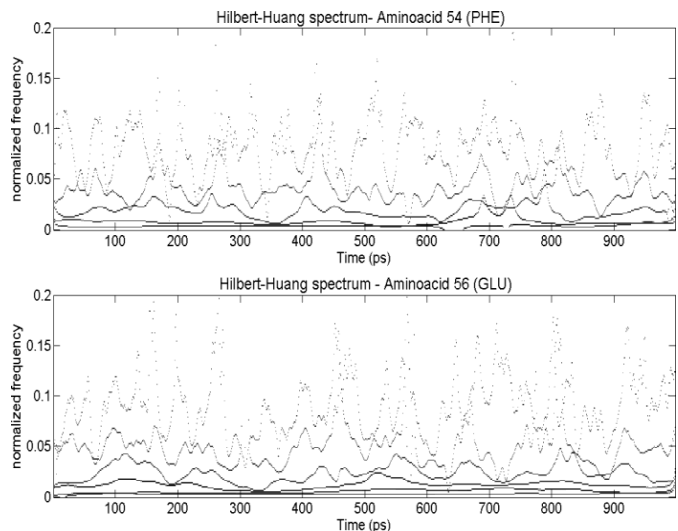

FIG. 5. Hilbert-Huang Spectrum of the $\psi$ trajectory for the residues 54 (PHE) and 56 (GLU).

In this work an attempt is made to understand the causes underlying the folding phenomena. The model proposed emphasizes that proteins adopt a three-dimensional structure through a dynamic process, which is expressed by cooperative movements of the peptide planes along the backbone. As the system evolves towards a more thermodynamically stable state, folding spontaneity is defined through the balance between entropic and enthalpic effects. However, it is important to consider that the nonlinear interaction between the peptide planes can give rise to a new emergent effect based on synchronization that has not yet been considered as part of the folding mechanism. In addition, this model also suggests that propagating conformational changes is associated with the transport of energy along the backbone as proposed by Botan et al [37]. In that sense, synchronization results from negative feedback with energy transfer between peptide planes when they are not properly oriented, which contribute to a wide energy distribution and led to a transition to a state with higher entropy. In the beginning of the process, amino acids behave independently, exploring a wide conformational space. Then hydrophobic forces drive the process and distant amino acids come together in space. Steric constraints limit the possible torsion angles that a residue can adopt. According to our hypothesis, the synchronization effect becomes more important as the mean interaction strength increases, which allows the formation of groups of synchronized elements. Such process is not uniform because it depends on interactions conditions that drive the system through various pathways. It is possible that many synchronized peptide planes pairs initially form. Subsequently, they group through a rapid aggregation process. Another possibility is that a large cluster of synchronized elements is formed first, and then it incorporates the additional peptide planes one by one. Synchronization in random and scale-free networks [38] follows similar pattern formations as described by different models of protein folding (DC and NC) [39]. The cooperativity is a key characteristic in two-state protein folding kinetics [40]. Many small proteins exhibit a two-state folding behavior without any accumulation of intermediates, going through a phase transition, from the unfolded to the native state. The self-synchronization transitions observed for coupled oscillator populations may explain this phenomenon. After that, noncovalent interactions help stabilize the native structure. Folding for large proteins, with more than 100 amino acids, is different from the previously described process because it involves the formation and modification of several clusters of synchronized peptide planes before the final transition into the native state. Based on the Kuramoto model (Eq. (3)), the stimulus for synchronization over a specific peptide plane (amino acid pair) is not pre-determined because it arises from the interactions between the set of amino acids, constituting a typical case of self-organization. Thus, the native protein structure forms due to the mean-field forces acting on a specific amino acid sequence. The same forces may explain why some sequence segments can assume either helix or sheet conformations in different proteins [41]. Defining such forces can improve the prediction of the protein structure from sequence alone. In that sense, equivalent forces generated by homologous proteins, which evolved from a common ancestor, may explain the success of homology structural modeling [42]. Likewise, changes in the composition of the forces caused, for example, by binding may explain the protein conformational changes seen in distant parts of the polypeptide chain, associated with the allosteric effect [43]. Interestingly, synchronization may also play a key role in the mechanism behind the sense of smell. Models based on shape [44] or vibrations [45, 46] cannot predict the response to the interaction between odorant molecules and protein receptors. In contrast to those models, synchronization as signal transduction overlaps both domains to explain the ligand-receptor selectivity. Thus, distinct odorants are recognized by different combinations of ligand-binding receptors according to their modified patterns of synchronization. This effect activates several receptors types and causes neurons to fire which determines the unique scent interpreted by the brain.

Finally, in conclusion, we further expected that thoroughly understanding synchronization may facilitate control over the folding process and help to predict a protein's native structure from its amino acid sequence.